\documentclass[11pt]{article}

\usepackage[latin1]{inputenc}
\usepackage[T1]{fontenc}
\usepackage[english]{babel}
\usepackage{amsfonts}
\usepackage{amssymb}
\usepackage{amsmath}
\usepackage{amsthm}
\usepackage{graphicx}
\def\be{\begin{equation}}
\def\ee{\end{equation}}
\def\bea{\begin{eqnarray}}
\def\eea{\end{eqnarray}}

\begin{document}

\title{\textbf{Replica symmetry breaking
in mean field spin glasses trough Hamilton-Jacobi technique}}

\author{Adriano Barra \footnote{Dipartimento di Fisica,
   Sapienza Universit\`a di Roma} \ Aldo Di Biasio\footnote{Dipartimento di Fisica, Universit\`a di Parma} \ Francesco Guerra\footnote{Dipartimento di Fisica,
   Sapienza Universit\`a di Roma \& Istituto Nazionale di Fisica Nucleare, Sezione di Roma $1$}}

%\date{March 2010}

\maketitle

\begin{abstract}
During the last years, through the combined effort of the insight, coming from physical intuition and computer simulation, and the exploitation of rigorous mathematical methods, the main features of the mean field Sherrington-Kirkpatrick spin glass model have been firmly established. In particular, it has been possible to prove the existence and uniqueness of the infinite volume limit for the free energy, and its Parisi expression, in terms of a variational principle, involving a functional order parameter. Even the expected property of ultrametricity, for the infinite volume states, seems to be near to a complete proof.
\newline
The main structural feature of this model, and related models, is the deep phenomenon of spontaneous replica symmetry breaking
(RSB), discovered by Parisi many years ago. By expanding on our previous work, the aim of this paper is to investigate a general frame, where replica symmetry breaking is embedded in a kind of mechanical scheme of the Hamilton-Jacobi type. Here, the analog of the ``time'' variable is a parameter characterizing the strength of the interaction, while the ``space'' variables rule out quantitatively the broken replica symmetry pattern. Starting from the simple cases, where annealing is assumed, or replica symmetry, we build up a progression of dynamical systems, with an increasing number of space variables, which allow to weaken the effect of the potential in the Hamilton-Jacobi equation, as the level of symmetry braking is increased.
\newline
This new machinery allows to work out mechanically the general
$K$-step RSB solutions, in a different interpretation with respect to the replica trick, and lightens easily their properties as existence or uniqueness.
\end{abstract}

\section{Introduction}

In the past twenty years the statistical mechanics of disordered
systems earned an always increasing weight as a powerful framework
by which analyze the world of complexity \cite{bara} \cite{amit}
\cite{bouchaud} \cite{CG} \cite{MPV} \cite{science} \cite{parisi}.
\newline
The basic model  of this field of research is the
Sherrington-Kirkpatrick model \cite{sk} (SK) for a spin glass, on which several
method of investigation have been tested along these years
\cite{alr} \cite{ass} \cite{barra1} \cite{bovier} \cite{contucci}
\cite{CLT} \cite{quadratic} \cite{talaRSB} \cite{talaHT}. The
first method developed has been the {\itshape replica trick}
\cite{sk2}\cite{parisi2} which, in a nutshell, consists in
expressing the quenched average of the logarithm of the partition function $Z(\beta)$ in the form $\mathbb{E}\ln Z(\beta) = \lim_{n\rightarrow 0}\mathbb{E}(Z(\beta)^n-1)/n$. Since the averages are easily calculated for integer values of $n$, the problem is to find the right analytic continuation allowing, in some way, to evaluate the $n \rightarrow 0$ limit, at the least for the case of large systems. Such analytic continuation is extremely complex, and many efforts have
been necessary to examine this  problem in the light
of theoretical physics tools, such as symmetries and their breaking
\cite{parisi3}\cite{parisi4}. In this scenario a solution has been
proposed by Parisi, with the well known Replica Symmetry Breaking
scheme (RSB), both solving the SK-model by showing a peculiar
``picture'' of the organization of the underlaying microstructure
of this complex system \cite{MPV}, as well as conferring a key
role to the replica-trick method itself \cite{challenge}.
\newline
The physical relevance, and deep beauty, of the results, obtained in the frame of the replica trick, have prompted a wealth of further research, in particular toward the objective of developing rigorous mathematical tool for the study of these problems. Let us recall, very schematically, some of the results obtained along these lines. 
Ergodic behavior has been
confirmed in \cite{comets}\cite{CLT}, the lack of 
self-average for the order parameter has been shown in
\cite{pastur}, the existence of the thermodynamic limit in
\cite{limterm}, the universality with respect to coupling's
distribution in \cite{carmona}, the correctness of the Parisi
expression for the free energy in 
\cite{g3}\cite{t4}, the critical behavior in \cite{barra3}, the
constraints to the free overlap fluctuations in
\cite{ac}\cite{gg}, and so much other contributions developed to
give rise even to textbooks (see for instance
\cite{bovierbook}\cite{hertz}\cite{challenge}).
\newline
Very recently, new investigations on ultrametricity started
(\cite{arguin}\cite{arguin2}) and allowed even  strong statements
dealing with the latter \cite{panchenko}, highlighting  as a
consequence the enquiry for techniques to prove the uniqueness of
the Parisi solution, step by step.
\newline
In this paper we match two other techniques, the broken replica
symmetry bound \cite{g3} and the Hamilton-Jacobi method
\cite{sum-rules}\cite{barra2}\cite{io1}, so to obtain a unified
and  stronger mathematical tool to work out free energies at
various levels of RSB, whose properties are easily available as
consequences of simple analogies with purely mechanical systems
\cite{io1}. We stress that within this framework, the improvement
of the free energy by increasing the replica symmetry breaking
steps is transparent.
\newline
In this first paper we show the method in full details, and
pedagogically apply it for recovering the annealed  and the
replica symmetric solutions, then we work out the first level of
RSB and show how to obtain the $1$-RSB Parisi solution with its
properties.
\newline
The paper is organized as follows: In Section ($2$) the SK model
is introduced together with its related statistical mechanics
definitions. In Section ($3$) the Broken Replica Mechanical
Analogy is outlined in full details (minor calculations are
reported in the Appendix), while Sections ($4,5,6$) are
respectively dedicated to the annealed, the replica symmetric and
the $1$-RSB solutions of the SK model with our approach. Section
($7$) deals with the properties of the solutions and Section ($8$)
is left for outlooks and conclusions.

\section{The Sherrington-Kirkpatrick mean field spin glass}

The generic configuration of the Sherrington-Kirkpatrick model
\cite{sk, sk2} is determined by the $N$ Ising variables
$\sigma_i=\pm1$, $i=1,2,\ldots,N$. The Hamiltonian of the model,
in some external magnetic field $h$, is
\begin{equation}
\label{SK} H_N(\sigma,h;J)=-\frac1{\sqrt N} \sum_{1 \leq i < j
\leq N} J_{ij} \sigma_i\sigma_j- h\sum_{1 \leq i \leq N} \sigma_i.
\end{equation}
The first term in (\ref{SK}) is a long range random two body
interaction, while the second represents the interaction of the
spins with the magnetic field $h$. The external quenched disorder
is given by the $N(N-1)/2$ independent and identically distributed
random variables $J_{ij}$, defined for each pair of sites. For the
sake of simplicity, denoting the average over this disorder by
$\mathbb{E}$, we assume each $J_{ij}$ to be a centered unit
Gaussian with averages
$$\mathbb{E}(J_{ij})=0,\quad \mathbb{E}(J_{ij}^2)=1.$$

For a given inverse temperature\footnote{Here and in the
following, we set the Boltzmann constant $k_{\rm B}$ equal to one,
so that $\beta=1/(k_{\rm B} T)=1/T$.} $\beta$, we introduce the
disorder dependent partition function $Z_{N}(\beta,h;J)$, the
quenched average of the free energy per site $f_{N}(\beta,h)$, the
associated averaged normalized log-partition function
$\alpha_N(\beta,h)$, and the disorder dependent Boltzmann-Gibbs
state $\omega$, according to the definitions
\begin{eqnarray}
\label{Z}
Z_N(\beta,h;J)&=&\sum_{\sigma}\exp(-\beta H_N(\sigma,h; J)),\\
\label{f}
-\beta f_N(\beta,h)&=&\frac1N \mathbb{E}\ln Z_N(\beta,h)=\alpha_N(\beta,h),\\
\label{state}
\omega(A)&=&Z_N(\beta,h;J)^{-1}\sum_{\sigma}A(\sigma)\exp(-\beta
H_N(\sigma,h;J)),
\end{eqnarray}
where $A$ is a generic function of $\sigma$.

Let us now introduce the important concept of replicas. Consider a
generic number $n$ of independent copies of the system,
characterized by the spin configurations $\sigma^{(1)}, \ldots ,
\sigma^{(n)}$, distributed according to the product state
%\begin{equation*}
$$
\Omega=\omega^{(1)} \times \omega^{(2)} \times \dots \times
 \omega^{(n)},
$$
%\end{equation*}
where each $\omega^{(\alpha)}$ acts on the corresponding
$\sigma^{(\alpha)}_i$ variables, and all are subject to the {\em
same} sample $J$ of the external disorder. These copies of the
system are usually called {\em real replicas}, to distinguish them
from those appearing in the  {\em replica trick} \cite{MPV}, which
requires a limit towards zero number of replicas ($n\to0$) at some
stage.

The overlap  between two replicas $a,b$ is defined according to
%\begin{equation}\label{q}
\begin{equation}
\label{overlap} q_{ab}(\sigma^{(a)},\sigma^{(b)})={1\over N}
\sum_{1 \leq i \leq N}\sigma^{(a)}_i\sigma^{(b)}_i,
\end{equation}
%\end{equation}
and satisfies the obvious bounds
%\begin{equation}
$$-1\le q_{ab}\le 1.$$
%\end{equation}
For a generic smooth function $A$ of the spin configurations on
the $n$ replicas, we define the averages $\langle A \rangle$ as
\begin{equation}\label{medie}
\langle A \rangle = \mathbb{E}\Omega
A\left(\sigma^{(1)},\sigma^{(2)},\ldots,\sigma^{(n)}\right),
\end{equation}
where the Boltzmann-Gibbs average $\Omega$ acts on the replicated
$\sigma$ variables and $\mathbb{E}$ denotes, as usual, the average
with respect to the quenched disorder $J$.

\section{Thermodynamics through a broken replica
mechanical analogy}

Once introduced the model, let us briefly discuss the plan we are
going to follow.
\newline
In the broken replica symmetry bound (BRSB) \cite{g3} it has been
shown that the Parisi solution is a bound for the true free energy
(the opposite bound has been achieved in \cite{t4}). This has been
done by introducing a suitable recursive interpolating scheme that
we are going to recall hereafter.
\newline
In the Hamilton-Jacobi technique instead \cite{sum-rules}, it has
been shown, by introducing a simple two parameter interpolating
function, how to recover the replica symmetric solution trhough a
mechanical analogy, offering as a sideline a simple prescription,
once the bridge to mechanics was achieved, to proof the uniqueness
of the replica symmetric solution.
\newline
The main result of this paper is that the two approaches can be
merged such that even the recursive interpolating structure of the
BRSB obeys a particular Hamilton-Jacobi description. This result
has both  theoretical and  practical advantages: the former is a
clear bridge among  improving approximation of the free energy
solution
 and increasing the levels of RSB, the latter is a completely
 autonomous  mechanical tool by which obtain solutions at various RSB
 steps in further models.
\newline
The task is however not trivial: the motion is no longer on a
$1+1$  Euclidean space-time as in \cite{sum-rules} but lives in
$K+1$ dimensions such that momenta and mass matrix need to be
introduced.

To start showing the whole procedure, let us introduce the
following \emph{Boltzmannfaktor}
\begin{equation}\label{bfaktor}
B(\{\sigma\} ; \bold{x},t) = \exp \left( \sqrt{\frac{t}{N}}
\sum_{(ij)} J_{ij} \sigma_i \sigma_j + \sum_{a=1}^{K} \sqrt{x_a}
\sum_i J_i^a \sigma_i \right)
\end{equation}
where both the $J_{ij}$s and the $J_i^a$s are standard Gaussian
random variables $\mathcal{N}[0,1]$ i.i.d. The $t$ parameter and
each of the $x_a$ maybe tuned in $\mathbb{R}^+$. We will use both
the symbol $\bold{x}$ as well as $(x_1,...,x_K)$ to label the $K$
interpolating real parameters coupling the one body interactions.
$K$ represents the dimensions, corresponding to the RSB steps in
replica trick. Let us denote via $E_a$ each of the averages with
respect to each of the $J_a$'s and $E_0$ the one with respect to
the whole $J_{ij}$ random couplings. Through eq. (\ref{bfaktor})
we are allowed to define the following partition function
$\tilde{Z}_N(t; x_1, \dots, x_K)$ and, iteratively, all the other
BRSB approximating functions for $a = 0, \dots, K$:
            \begin{eqnarray}
                Z_K & \equiv & \tilde{Z}_N = \sum_{\sigma}B(\{ \sigma \};\bold{x},t), \\
                & \dots & \nonumber \\
                Z_{a-1}^{m_a} & \equiv & E_a \left( Z_{a}^{m_a} \right), \\
                & \dots & \nonumber \\
                Z_0^{m_1} & \equiv & E_1 \left( Z_{1}^{m_1} \right).
            \end{eqnarray}
            We need further to introduce the following
            interpolating function
            \begin{equation}
            \label{alfa}
                \tilde{\alpha}_N(t; x_1, \dots, x_K) \equiv \frac{1}{N} E_0 \log
                Z_0,
            \end{equation}
 and define, for $a=1, \dots, K$, the random variables
            \begin{equation}
                f_a \equiv \frac{Z_a^{m_a}}{E_a \left( Z_a^{m_a}
                \right)},
            \end{equation}
   and the generalized states
            \begin{equation}
                \tilde{\omega}_a(.) \equiv E_{a+1} \dots E_K \left( f_{a+1} \dots f_K \omega (.)
                \right),
            \end{equation}
          the whole in complete analogy with the "broken prescriptions'' \cite{g3}.
\newline
            Of course the corresponding replicated states $\Omega_a$ are immediately generalized
            with respect to each of the $\omega_a$ state introduced above.

            Overall, for $a=0, \dots, K$, we further need the averages
            \begin{equation}
                \langle.\rangle_a \equiv E \left( f_{1} \dots f_a \tilde{\Omega}_a (.) \right).
            \end{equation}

While it is clear that, when evaluated at $t=\beta^2$ and
$\bold{x}=0$, our interpolating function
$\tilde{\alpha}(t,\bold{x})$ reproduces the definition of the
quenched free energy,  when evaluated at $t=0$ (which a proper
choice for the $\bold{x}$ parameters that we are going to show),
it reproduces the Parisi trial solution $f(q=0, y=h)$ at the given
$K$ level of RSB:
                \begin{eqnarray}
                    & & \tilde{\alpha}_N (t=0; x_1, \dots, x_K) = \nonumber \\
                    & & = \frac{1}{N} \log
                    \left[
                    E_1 \dots
                    \left[
                    E_K \left( \sum_{\sigma} \exp \left(\sum_{a=1}^K \sqrt{x_a} \sum_i J_i^a \sigma_i \right) \right)^{m_K}
                    \right]^{\frac{1}{m_K}}
                    \dots \right]^{\frac{1}{m_1}}.
                \end{eqnarray}
Even though far from being trivial, this is an essential feature
of mean field behavior even in the disordered framework; in fact,
in the thermodynamic limit the connected correlation inside pure
states should go to zero bridging the two body problem to a
(collection of) one body model, or better "high temperature
model'', whose partition function factorizes:
                \begin{equation}
                    \sum_{\sigma} \exp \left(\sum_{a=1}^K \sqrt{x_a} \sum_i J_i^a \sigma_i \right) =
                    2^N \prod_i \cosh \left( \sum_{a=1}^K \sqrt{x_a} J_i^a
                    \right),
                \end{equation}
such that, averaging over the $J_i^K$, we get
                \begin{eqnarray}
                    &&E_K \left(
                    \sum_{\sigma} \exp \left(
                    \sum_{a=1}^K \sqrt{x_a} \sum_i J_i^a \sigma_i
                    \right)
                    \right)^{m_K} = \nonumber \\
                    &&=
                    2^{N m_K} \prod_i
                    \int d\mu (z_K) \cosh^{m_K} \left(
                    \sum_{a=1}^{K-1} \sqrt{x_a} J_i^a + z_K \sqrt{x_K}
                    \right),
                \end{eqnarray}
and so on. Even taking the external field $h$, which is again
encoded in a single body interaction and is simply added into the
hyperbolic cosine, we get
                \begin{eqnarray}
                    \tilde{\alpha}_N (t&=&0; x_1, \dots, x_K) = \log 2 + \nonumber \\
                    &+&
                    \log \left[
                    \int d\mu(z_1) \dots
                    \left[
                    \int d\mu(z_K)
                    \cosh^{m_K} \left(\sum_{a=1}^K \sqrt{x_a} z_a + \beta h \right)
                    \right]^{\frac{1}{m_K}}
                    \dots \right]^{\frac{1}{m_1}}. \nonumber
                \end{eqnarray}
                In the case where
                $
                x_a = \beta^2 (q_a - q_{a-1})
                $
                the second term does coincide sharply with the solution of the Parisi equation \cite{MPV}.

Let us now define $S(t,\bold{x})$ as Principal Hamilton Function
(PHF) for our problem:
                \begin{equation}
                    S(t; x_1, \dots, x_K) =
                    2 \left(
                    \alpha (t; x_1, \dots, x_K) - \frac{1}{2} \sum_{a=1}^K x_a - \frac{1}{4}t
                    \right).
                \end{equation}
As proved in the Appendix, the $(x,t)$-streaming of $S(t; x_1,
\dots, x_K)$ are then
                \begin{eqnarray}
                \label{dts}
                    \partial_t S(t; x_1, \dots, x_K) & = &
                    -\frac{1}{2} \sum_{a=0}^K (m_{a+1} - m_a) \langle q_{12}^2 \rangle_a, \\
                \label{das}
                    \partial_a S(t; x_1, \dots, x_K) & = &
                    -\frac{1}{2} \sum_{b=a}^K (m_{b+1} - m_b) \langle q_{12}
                    \rangle_b.
                \end{eqnarray}
It is then possible to introduce an Hamilton-Jacobi structure for
$S(x,t)$, which implicity defines a potential $V(t; x_1, \dots,
x_K)$, so to write
                \begin{equation}
                \label{hj}
                    \partial_t S(t,\bold{x}) + \frac{1}{2} \sum_{a,b=1}^K \partial_a S \, (M^{-1})_{ab} \, \partial_b S + V(t,\bold{x}) = 0.
                \end{equation}
The kinetic term reads off as
                \begin{eqnarray}
                    T & \equiv & \frac{1}{2} \sum_{a,b=1}^K \partial_a S (t; x_1, \dots, x_K) \, (M^{-1})_{ab} \, \partial_b S (t; x_1, \dots, x_K) \nonumber \\
                    &=& \frac{1}{2} \sum_{a,b=1}^K (M^{-1})_{ab} \,
                    \sum_{c \geq a}^K \sum_{d \geq b}^K
                    (m_{c+1} - m_c) \langle q_{12} \rangle_c (m_{d+1} - m_d) \langle q_{12} \rangle_d \nonumber \\
                    &=& \frac{1}{2} \sum_{c,d=1}^K D_{cd} \,
                    (m_{c+1} - m_c) \langle q_{12} \rangle_c (m_{d+1} - m_d) \langle q_{12} \rangle_d, \nonumber \\
                \end{eqnarray}
where we defined
                \begin{equation}
                    D_{cd} \equiv \sum_{a=1}^c \sum_{b=1}^d (M^{-1})_{ab}.
                \end{equation}
By the inversion of the mass matrix
                \begin{equation}
                \label{dcdm}
                    D_{cd} (m_{c+1} - m_c) = \delta_{cd}
                \end{equation}
we obtain the expression
                \begin{eqnarray}
                    T & = & \frac{1}{2} \sum_{c=1}^K (m_{c+1}- m_c) \langle q_{12} \rangle_c^2 \\
                    \label{kin}
                    &  = & \frac{1}{2} \sum_{c=0}^K (m_{c+1}- m_c) \langle q_{12} \rangle_c^2
                    - \frac{1}{2} (m_1 - m_0) \langle q_{12}
                    \rangle_0^2.
                \end{eqnarray}

Condition (\ref{dcdm}) determines the elements of the inverse of
the mass matrix $M^{-1}$. \newline In particular we stress that it
is symmetric and the not zero values are only on the diagonal and
all  of them respecting $(M^{-1})_{a, a+1} = (M^{-1})_{a+1, a}$:
                \begin{eqnarray}
                    (M^{-1})_{11} & = & \frac{1}{m_2-m_1}, \\
                    \label{invmaa}
                    (M^{-1})_{a, a} & = & \frac{1}{m_{a+1}-m_a} + \frac{1}{m_a-m_{a-1}}, \\
                    (M^{-1})_{a, a +1} & = & -
                    \frac{1}{m_{a+1}-m_a},
                \end{eqnarray}
all the others being zero.
\newline
%Without loosing generality, for the last diagonal term
%$(M^{-1})_{KK}$, we can impose $m_K < 1$ such that
%eq.(\ref{invmaa}) continues to hold.
The elements of the mass matrix $M$ are determined by the equation
                \begin{equation}
                    \sum_b M_{ab} (M^{-1})_{bc} = \delta_{ac},
                \end{equation}
and it is immediate to verify that the following representation
holds:
                \begin{equation}
                    M_{ab} = 1 - m_{(a \wedge b)}.
                \end{equation}
With this expression for the matrix elements, by substituting eq.s
(\ref{dts}) and (\ref{kin}) into (\ref{hj}) we obtain the
expression for the potential such that overall
\begin{eqnarray}
\partial_t S(t; x_1, \dots, x_K) &+& \frac{1}{2} \sum_{a,b=1}^K \partial_a S \, (M^{-1})_{ab} \, \partial_b S + V(t; x_1, \dots, x_K) = 0, \\
V(t; x_1, \dots, x_K) &=& \frac{1}{2} \sum_{a=0}^K (m_{a+1} - m_a)
                    (\langle q_{12}^2 \rangle_a - \langle q_{12} \rangle_a^2)
                    + \frac{1}{2} (m_1 - m_0) \langle q_{12}
                    \rangle_0^2. \nonumber
                \end{eqnarray}
Once the mechanical analogy is built, it is however prohibitive
solving the problem as it is (i.e. integrate the equations of
motion); instead we propose an iterative scheme that mirrors the
replica symmetry breaking one: at first, by choosing $K=1$, we
solve the free field solution (we impose $V(t,\bold{x})=0$) and we
recover the annealed expression for the free energy. This is
consistent with neglecting the potential as it turns out to be the
squared overlap. Then, we avoid perturbation scheme to deal with
the source but we enlarge our Euclidean space by considering
$K=2$. Again we work out the free field solution to obtain the
replica-symmetric expression for the free energy, consistently
with neglecting the potential; in fact the source we avoid, this
time, is the variance of the overlap: a much better approximation
with respect to $K=1$. We go further explicitly by considering the
$K=3$ case and we get the $1$-RSB solution in the same way (and so
on). Interesting we discover that there is a one to one connection
among the steps of replica symmetry breaking in replica trick and
the Euclidean dimension  in the broken replica mechanical analogy.
The latter however incorporates, in a single scheme, even the
annealed and the replica symmetry solutions.

%%%%%%%%%%%%%%%%%%%%%%%%%%%%%%%%%%%%% (K=1) ANNEALING %%%%%%%%%%%%%%%%%%%%%%%%%%%%%%%%%%%%%%%%%%%%%%%%%%%%%%%%%%%%%
        \section{$K=1$, Annealed free energy}

Let us now recover some properties of disordered thermodynamics by
studying the $K=1$ case so to show how the solution of the free
problem coincides with the annealed expression.
\newline
We assume $x(q)=m_1=1$ in the whole interval $[0,1]$.
\newline
We show now that, within our approach, this implies a reduction in
the degrees of freedom where the Hamilton-Jacobi action lives,
such that the PHF depends by $t$ only.

The dynamics involves a $1+1$ Euclidean space-time such that
            \begin{equation}
                Z_1 \equiv Z_K \equiv \tilde{Z}_N \equiv
                \sum_{\sigma} \exp \left(
                \sqrt{t/N}H_N(\sigma;J) + \sqrt{x} \sum_i J_i \sigma_i \right).
            \end{equation}
            $Z_0$ is consequently given by
            \begin{equation}
                Z_0 \equiv E_1 Z_1 =
                \exp \left(\frac{N}{2}x \right)
                \sum_{\sigma} \exp \left(\sqrt{t/N}H_N(\sigma;J) \right).
            \end{equation}
            This implies, into the interpolating function, a
            linear and separate dependence by the $x$
            \begin{equation}
                \tilde{\alpha}(t,x) = \frac{x}{2}
                + \frac{1}{N} E_0 \log \sum_{\sigma} \exp
                \left( \sqrt{t/N}H_N(\sigma;J)\right).
            \end{equation}
The $x$-derivative of $\tilde{\alpha}(t,\bold{x})$ is immediate,
while for the $t$-one we can use the general expression previously
obtained (cfr. eq.s (\ref{dts},\ref{das}))
            \begin{eqnarray}
            \partial_t \tilde{\alpha} & = & \frac{1}{4}
                \left[ 1 - \langle q_{12}^2 \rangle_0 \right],\\
                \partial_x \tilde{\alpha} & = & \frac{1}{2}.
            \end{eqnarray}
    As a straightforward but interesting consequence,
    PHF does not depend on $x$ and we get
            \begin{eqnarray}
                S(t,x) & = &  2 \tilde{\alpha}(t,x) - x - \frac{t}{2} \nonumber \\
                & = & \frac{2}{N} E_0 \log \sum_{\sigma} \exp
                \left( \sqrt{t/N}H_N(\sigma;J) \right)
                - \frac{t}{2}, \\
                \partial_t S & = & - \frac{1}{2} \langle q_{12}^2
                \rangle_0,\\
\partial_x S & \equiv & v(t,x) = 0, \\
            \end{eqnarray}
where $v(t)$ defines the velocity field, which is identically zero
such that $ x(t) \equiv x_0 $.
\newline
In this simplest case, the potential is trivially the
$t$-derivative of  $S(t,\bold{x})$ with a change in the sign, (the
averaged squared overlap):
            \begin{equation}
                V(t,\bold{x}) = \frac{1}{2} \langle q_{12}^2 \rangle_0.
            \end{equation}

Now we want to deal with the solution of the statistical mechanics
problem. As we neglect the source (we are imposing $\langle
q_{12}^2 \rangle_0=0$), we can take the initial value for $S(x,t)$
as it must be constant overall the space-time.
            \begin{equation}
                \bar{S} = S (0) = 2 \log 2,
            \end{equation}
and, consequently, we can write the solution of the problem as
            \begin{equation}
                \bar{\alpha}(t, x) = \log 2 + \frac{x}{2} + \frac{t}{4}.
            \end{equation}
At this point it is straightforward to obtain statistical
mechanics by posing $t = \beta^2$ and $x = 0$:
            \begin{equation}
                {\alpha}_N(\beta) = \log 2 + \frac{\beta^2}{4},
            \end{equation}
            which is exactly the annealed free energy.

%%%%%%%%%%%%%%%%%%%%%%%%%%%%%%%%%%%%%% (K=2) REPLICA SIMMETRICA %%%%%%%%%%%%%%%%%%%%%%%%%%%%%%%%%%%%%%%%%%%%%

        \section{$K=2$, Replica symmetric free energy}

In this section, by adding another degree of freedom to our
mechanical analogy, we want to reproduce the replica symmetric
solution of the statistical mechanics problem.
\newline
We deal with $K=2$. The order parameter is now taken as
            \begin{equation}
                \label{xqbar}
                x(q) = x_{\bar{q}}(q) =
                \left\{
                \begin{array}{ll}
                    0 & \mbox{if } q \in [0, \bar{q}), \\
                  1 & \mbox{if } q \in [\bar{q}, 1].
                \end{array}
                \right.
            \end{equation}
            So\begin{center}
            \begin{eqnarray}
                & & q_1 = \bar{q}, \quad q_2 = q_K \equiv 1 \\
                & & m_0 = m_1 = 0, \quad m_2 = m_K = 1, \quad m_3 = m_{K+1} \equiv 1.
            \end{eqnarray}\end{center}

The  auxiliary partition function depends on $t$ and on the two
spatial coordinates $x_1$ and $x_2$:
            \begin{equation}
                \tilde{Z}_N (t; x_1, x_2) \equiv
                \sum_{\sigma} \exp \left(\sqrt{t/N}H_N(\sigma;J)
                + \sqrt{x_1} \sum_i J_i^1 \sigma_i + \sqrt{x_2} \sum_i J_i^2
                \sigma_i\right),
            \end{equation}
            and with the latter, recursively, we obtain $Z_0$.
            \begin{eqnarray}
                Z_K & \equiv & Z_2 \equiv \tilde{Z}_N, \\
                Z_1 & \equiv & \left(E_2 Z_2^{m_2} \right)^{\frac{1}{m_2}} = E_2 Z_2, \\
                Z_0 & = &  \left(E_1 Z_1^{m_1} \right)^{\frac{1}{m_1}}.
            \end{eqnarray}
            The function $Z_1$ can be immediately evaluated by
            standard Gaussian integration as
            \begin{equation}
                Z_1 = \exp \left( N \frac{x_2}{2} \right)
                \sum_{\sigma} \exp \left(
               \sqrt{t/N}H_N(\sigma;J)
                + \sqrt{x_1} \sum_i J_i^1 \sigma_i \right).
            \end{equation}
Concerning the function $Z_0$ we can write
            \begin{eqnarray*}
                \left(E_1 Z_1^{m_1} \right)^{\frac{1}{m_1}} & = & \exp \left[
                \frac{1}{m_1} \log E_1 \left[
                \exp \left( m_1 \log Z_1 \right)
                \right] \right] \\
                & = & \exp \left[
                \frac{1}{m_1} \log E_1 \left[
                1 + m_1 \log Z_1 + o (m_1^2)
                \right] \right] \\
                & = & \exp \left[
                \frac{1}{m_1} \left[
                m_1 E_1 \log Z_1 + o (m_1^2)
                \right] \right] \\
                & = & \exp E_1 \log Z_1 + o (m_1),
            \end{eqnarray*}
            and consequently
            \begin{equation}
                Z_0 = \exp E_1 \log Z_1.
            \end{equation}
In this case, our interpolating function reads off as
            \begin{equation}
            \label{alfars}
                \tilde{\alpha}(t, x_1, x_2) = \frac{x_2}{2} +
                \frac{1}{N} E_0 E_1 \log \left[
                \sum_{\sigma} \exp \left(\sqrt{t/N}H_N(\sigma;J)
                + \sqrt{x_1} \sum_i J_i^1 \sigma_i  \right)
                \right].
            \end{equation}
Again by using the general formulas sketched in the first section
(cfr. eq.s (\ref{dts},\ref{das})) we get for the derivatives
            \begin{eqnarray}
                \partial_t \tilde{\alpha} & = & \frac{1}{4} \left[1 - \langle q_{12}^2 \rangle_1 \right], \\
                \partial_{x_1} \tilde{\alpha} & = & \frac{1}{2} \left[1 - \langle q_{12} \rangle_1 \right], \\
                \partial_{x_2} \tilde{\alpha} & = & \frac{1}{2}.
            \end{eqnarray}
Evaluating our function at $t=0, x_1=x_1^0, x_2=x_2^0$ we easily
find
            \begin{equation}
                \tilde{\alpha}(0; x_1^0, x_2^0) =
                \frac{x_2^0}{2}
                + \log 2
                + \int d \mu(z) \, \log \cosh (\sqrt{x_1^0} \, z).
            \end{equation}

Let us introduce now the $K=2$ PHF
            \begin{equation}
                S(t; x_1, x_2) = 2 \left( \tilde{\alpha} - \frac{x_1}{2}
                - \frac{x_2}{2}- \frac{t}{4} \right),
            \end{equation}
together with its  derivatives
            \begin{eqnarray}
                \partial_t S & = & - \frac{1}{2} \langle q_{12}^2 \rangle_1, \\
                \partial_{x_1} S & = & v_1 (t, x_1) = - \langle q_{12} \rangle_1, \\
                \partial_{x_2} S & = & 0.
            \end{eqnarray}
We observe that, even in this case, there is no true dependence by
one of the spatial variables ($x_2$): this is due to the constant
value of the last interval  $m_K = m_2$ where the order parameter
equals one and can be Gaussian-integrated out immediately into the
corresponding $Z_2$ getting the pre-factor $\exp(\frac{1}{2} N
x_2^0)$.
\newline
As a consequence, we can forget the mass matrix as there is no
true multidimensional space.
\newline
Let us write down the Hamilton-Jacobi equation
            \begin{equation}
                \partial_t S(t, x_1)
                + \frac{1}{2} \left( \partial_{x_1} S (t, x_1) \right)^2
                + V (t, x_1) = 0.
            \end{equation}
The potential is given by the function
            \begin{equation}
                V(t, x_1) = \frac{1}{2}
                \left( \langle q_{12}^2 \rangle_1 - \langle q_{12} \rangle_1^2
                \right),
            \end{equation}
            where
            \begin{equation}
                \langle q_{12}^2 \rangle_1 =
                E_0 E_1 f_1 \Omega_1 (q_{12}^2) =
                E_0 E_1 f_1 \frac{1}{N^2} \sum_{ij}
                \left( E_2 f_2 \omega (\sigma_i \sigma_j) \right)^2.
            \end{equation}
When taking $x_1 = 0$ and $t = \beta^2$ the variance of the
overlap becomes the source of the streaming.
            \begin{equation}
                V(\beta^2, 0) = \frac{1}{2}
                \left( \langle q_{12}^2 \rangle - \langle q_{12} \rangle^2 \right).
            \end{equation}

As usual in our framework, we kill the source (i.e.
$V(t,\bold{x})=0$), and obtain for the velocity
            \begin{equation}
                \bar{q}(x_1^0) \equiv - v_1(0, x_1^0)
                = \int d \mu (z) \, \tanh^2 (z \sqrt{x_1^0}).
            \end{equation}
This is the well known self-consistency relation of Sherrington
and Kirkpatrick, namely
            \begin{equation}
                \bar{q}(\beta) =
                \int d \mu (z) \, \tanh^2 (\beta \sqrt{\bar{q}}
                z).
            \end{equation}
The free field solution of the Hamilton-Jacobi equation is then
the solution in a particular point (and of course the choice is $
\bar{S} (0, x_1^0)$ which requires only a one-body evaluation)
plus the integral of the Lagrangian over the time (which is
trivially built by the kinetic term alone when considering free
propagation). Overall the solution reads off as
            \begin{equation}
                \bar{S} (t, x_1) =
                \bar{S} (0, x_1^0)
                + \frac{1}{2} \bar{q}^2(x_1^0) t,
            \end{equation}
            by which
            statistical mechanics is recovered as usual, obtaining for the pressure
            \begin{equation}
                \bar{\alpha} (t; x_1, x_2) =
                \log 2 + \int d \mu (z) \, \log \cosh (\sqrt{x_1^0}z)
                + \frac{t}{4} ( 1 - \bar{q} )^2
                + \frac{x_2}{2},
            \end{equation}
that corresponds exactly to the replica-symmetric solution once
evaluated at $x_1=x_2=0$ and $t=\beta^2$ and noticing that
$0=x(t)=x_1^0-\bar{q}t$.
\newline
Within our description it is not surprising that the replica
symmetric solution is a better description with respect to the
annealing. In fact, while annealing is obtained neglecting the
whole squared overlap $\langle q_{12}^2\rangle$ as a source term,
the replica symmetric solution is obtained when neglecting only
its variance.
\newline
Of course, nor the former neither the latter may correspond to the
true solution. However we are understanding that increasing the
Euclidean dimensions (the RSB steps in replica framework)
corresponds to lessening the potential in the Hamilton-Jacobi
framework, and consequently reducing the error of the free field
approximation toward the true solution.

\section{$K=3$, $1$-RSB free energy}

    The simplest expression of $x(q)$ which breaks replica
    symmetry is obtainable when considering $K=3$,
    \bea
        && 0 = q_0 < q_1 < q_2 < q_3 = 1, \\
        && 0 = m_1 < m_2 \equiv m < m_3 = 1.
    \eea
    With this choice for the parametrization of $x(q)$ the solution of the Parisi equation
    \be
        \partial_q f + \frac{1}{2} \partial^2_y f + \frac{1}{2} x (\partial_y f)^2 = 0
    \ee
    is given by
    \bea
        f(0, h; x, \beta) & = & \frac{1}{m} \int d\mu (z_1) \log \int d\mu (z_2)
        \cosh^m [ \beta (\sqrt{q_1} z_1 + \sqrt{q_2 - q_1} z_2 + h)] + \nonumber \\
        & + & \frac{1}{2} \beta^2 (1 - q_2),
    \eea
    and, using a label $P$ to empathize that we are considering Parisi prescription, the pressure becomes
    \bea \nonumber
        \alpha_P ( \beta, h; x) & = & \log 2 + f(0, h; x, \beta) - \frac{1}{2} \beta^2 \int_0^1 q \, x(q) \, dq \\
    \label{eq:alfa}
        & = & \log 2 - \frac{1}{4} \beta^2 [ (m-1) q_2^2 -1 - m q_1^2 + 2 q_2] +  \\
        & + & \frac{1}{m} \int d\mu (z_1) \log \int d\mu (z_2)
        \cosh^m [ \beta (\sqrt{q_1} z_1 + \sqrt{q_2 - q_1} z_2 + h)]. \nonumber
    \eea

Now we want to see how it is possible to obtain this solution by
analyzing the geodetics of our free mechanical propagation in
$3+1$ dimensions.

\bigskip

Let us define
    \be
        \tilde{Z}_N (t; x_1, x_2, x_3) \equiv \sum_{\sigma} \exp \left[
        \sqrt{t/N}H_N(\sigma;J) + \sum_{a=1}^3 \sqrt{x_a} \sum_i J_i^a \sigma_i
        \right],
    \ee
    by which
    \bea
        Z_3 & \equiv & Z_K \equiv \tilde{Z}_N, \\
        Z_2 & = & E_3 Z_3 = \exp \left(\frac{N x_3}{2} \right) \sum_{\sigma} \exp \left[
         \sqrt{t/N}H_N(\sigma;J) + \sum_{a=1}^2 \sqrt{x_a} \sum_i J_i^a \sigma_i \right], \\
        Z_1 & = & \left(E_2 Z_2^m \right)^{1/m}, \\
        Z_0 & = & (E_1 Z_1^{m_1})^{1/m_1} = \exp (E_1 \log Z_1)
        = \exp \left[ \frac{1}{m} E_1 \log E_2 Z_2^m \right].
    \eea
    For the interpolating function we get in this way
    \bea
        && \tilde{\alpha}_N (t; x_1, x_2, x_3) \equiv  \frac{1}{N} E_0 \log Z_0 = \\
       && =  \frac{x_3}{2} + \frac{1}{Nm} E_0 E_1 \log \left\{ E_2 \left[ \sum_{\sigma} \exp \left(
       \sqrt{t/N}H_N(\sigma;J) + \sum_{a=1}^2 \sqrt{x_a} \sum_i J_i^a \sigma_i \right) \right]^m \right\}, \nonumber
    \eea
while for the derivatives we can use the general formulas, so to
obtain
    \bea
        \partial_t \tilde{\alpha} & = & \frac{1}{4} [1 - m \langle q_{12}^2 \rangle_1
        - (1-m) \langle q_{12}^2 \rangle_2 ], \\
        \partial_1 \tilde{\alpha} & = & \frac{1}{2} [1 - m \langle q_{12} \rangle_1
        - (1-m) \langle q_{12} \rangle_2 ], \\
        \partial_2 \tilde{\alpha} & = & \frac{1}{2} [1 - (1-m) \langle q_{12} \rangle_2 ], \\
        \partial_3 \tilde{\alpha} & = & \frac{1}{2}.
    \eea
Then we need to evaluate the interpolating function at the
starting time:
    \bea
        \tilde{\alpha}_N (0; x_1^0, x_2^0, x_3^0) & = & \frac{x_3}{2} + \log2 +\\
        & + &\frac{1}{m} \int d\mu(z_1) \log \left[ \int d\mu (z_2)
        \cosh^m \left( \sqrt{x_1^0} z_1 + \sqrt{x_2^0} z_2) \right) \right]. \nonumber
    \eea

The $K=3$ PHF, as usual and previously explained for the $K=1,2$
cases, does not depend on the last coordinate (i.e. $x_3$), such
that we can ignore it when studying the properties of the
solution.
    \bea
        S(t; x_1, x_2) & = & \frac{2}{Nm} E_0 E_1 \log E_2 \left[ \sum_{\sigma} \exp \left(
        \sqrt{tN/2}K(\sigma) + \sum_{a=1}^2 \sqrt{x_a} \sum_i J_i^a \sigma_i \right) \right]^m
        \nonumber \\
        & - & x_1 - x_2 - t/2.
    \eea
and the derivatives, implicitly defining the momenta (labeled by
$p_1, p_2$), are given by
    \bea
        \partial_t S & = & - \frac{m}{2} \langle q_{12}^2 \rangle_1
        - \frac{1-m}{2} \langle q_{12}^2 \rangle_2, \\
        \partial_1 S & \equiv & p_1 (t; x_1, x_2) = -m \langle q_{12} \rangle_1
        -(1-m)\langle q_{12} \rangle_2, \\
        \partial_2 S & \equiv & p_2 (t; x_1, x_2) = -(1-m)\langle q_{12} \rangle_2.
    \eea
   The kinetic energy consequently turns out to be
    \be
        T = \frac{m}{2} \langle q_{12} \rangle_1^2
        + \frac{1-m}{2} \langle q_{12} \rangle_2^2,
    \ee
    and the potential, which we are going to neglect as usual, is
    given by
    \be
        V(t; x_1, x_2) = \frac{1}{2} \left[
        m \left( \langle q_{12}^2 \rangle_1 - \langle q_{12} \rangle_1^2 \right)
        + (1-m) \left( \langle q_{12}^2 \rangle_2 - \langle q_{12} \rangle_2^2 \right) \right].
    \ee
By having two spatial degrees of freedom, the mass matrix has a $2
\times 2$ structure now
    \be
    M^{-1} = \left(\begin{array}{cc}
            1/m & -1/m \\
            -1/m & 1/[m(1-m)]
        \end{array}\right),
    \ee
    \be
    M = \left(\begin{array}{cc}
            1 & 1-m \\
            1-m & 1-m
        \end{array}\right).
    \ee
    Note that the eigenvalues of the mass matrix are always
    positive defined for $m\in[0,1]$.
    \newline
   We can determine now the velocity field
    \bea
        v_1(t; x_1, x_2) & = & \sum_{b=1}^2 (M^{-1})_{1b} \, p_b = - \langle q_{12} \rangle_1, \\
        v_2(t; x_1, x_2) & = & \sum_{b=1}^2 (M^{-1})_{2b} \, p_b =
        \langle q_{12} \rangle_1 - \langle q_{12} \rangle_2.
    \eea

So we get all the ingredients for studying the free field solution
(the one we get neglecting the source). In this case the equations
of motion are
    \bea
        x_1 (t) & = & x_1^0 - \langle q_{12} \rangle_1(0; x_1^0, x_2^0) \, t
        \equiv x_1^0 - \bar{q}_1 t \\
        x_2 (t) & = & x_2^0 + \left( \bar{q}_1
        - \langle q_{12} \rangle_1(0; x_1^0, x_2^0) \right) \, t
        \equiv x_2^0 + (\bar{q}_1 - \bar{q}_2 ) t
    \eea
and we can see that $\bar{q}_1$ and $\bar{q}_2$ satisfy the
self-consistency relations in agreement with  the replica trick
predictions \bea
        \label{eq:q_1}
            q_1 & = & \int d\mu(z) \left[
            D^{-1}(z) \int d\mu (y)
            \cosh^m \theta(z, y)
            \tanh \theta(z, y)
            \right]^2, \\
        \label{eq:q_2}
            q_2 & = & \int d\mu(z) \left[
            D^{-1}(z) \int d\mu (y)
            \cosh^m \theta(z, y)
            \tanh^2 \theta(z, y)
            \right], \\
            \theta(z, y) & = & \beta (\sqrt{q_1} z + \sqrt{q_2 - q_1} y), \\
            D(z) & = & \int d\mu (y) \cosh^m \theta(z, y).
        \eea
\newline
The PHF is obtained in coherence with the previous cases and obeys
    \be
        \bar{S} (t; x_1, x_2) = \bar{S} (0; x_1^0, x_2^0)
        + T(0; x_1^0, x_2^0) t,
    \ee
by which
    \be
        \bar{\alpha}(t;x_1, x_2, x_3) - \frac{x_1}{2} - \frac{x_2}{2} - \frac{t}{4} =
        \bar{\alpha}(0;x_1^0, x_2^0, x_3^0) - \frac{x_1^0}{2} - \frac{x_2^0}{2} + T (0; x_1^0, x_2^0)
        \frac{t}{2},
    \ee
   and, remembering that
    \bea
        x_1-x_1^0 & = & -\bar{q}_1 t, \\
        x_2-x_2^0 & = & (\bar{q}_1-\bar{q}_2) t,
    \eea
we get the thermodynamic pressure in the space-time coordinates:
    \bea
        \bar{\alpha}(t;x_1, x_2, x_3) & = & \frac{x_3}{2} + \log 2
        - \frac{t}{4}[ -1 + 2 \bar{q}_2 - m \bar{q}_1^2 - (1-m) \bar{q}_2^2 ]  \\
        &+&  \frac{1}{m} \int d\mu(z_1) \log \left[ \int d\mu (z_2)
        \cosh^m \left[ \sqrt{x_1^0} z_1 + \sqrt{x_2^0} z_2) \right] \right]. \nonumber
    \eea
In order to get the statistical mechanics result, as usual, we
need to evaluate the latter in  $t = \beta^2$,
    $x_1=x_2=x_3=0$, from which $x_1^0 = \bar{q}_1 t$ and $x_2^0 = (\bar{q}_1 - \bar{q}_2) t$,
    gaining once again (\ref{eq:alfa}).

\section{Properties of the $K=1,2,3$ free energies}

In the previous sections, we obtained solutions for the
Hamilton-Jacobi equation in the $K=1, 2, 3$ cases, without saying
anything about uniqueness. For $K=1$, the annealed case, there is
no true motion so it is clear that there is just a single straight
trajectory, identified by the initial point $x_0 = x$,
intersecting the generic point $(x,t)$, with $x, t > 0$.

In the $K=2$ problem, well studied in \cite{sum-rules}, one can
show uniqueness by observing that the function $t(x_0)$,
representing the point at which the trajectory intersects the
$x$-axis, is a monotone increasing one of the initial point $x_0$,
so that given $x, t > 0 $, there is a unique point $x_0$ (and
velocity $\bar{q}(x_0)$, of course) from which the trajectory
starts.

For $K=3$, the problem becomes much complicated, because we now
have to consider motion in a three dimensional Euclidean space,
proving that given the generic point $(x_1, x_2, t)$, with
$x_1>0$, $x_2>0$, $t>0$, there exists a unique line passing in
$(x_1, x_2)$ at time $t$.

So let us consider the functions \bea
    F(x_1, t; x_1^0, x_2^0) & \equiv & x_1 - x_1^0 + \bar{q}_1(x_1^0, x_2^0) t, \\
    G(x_2, t; x_1^0, x_2^0) & \equiv & x_2 - x_2^0 + \bar{q}_2(x_1^0, x_2^0) t - \bar{q}_1(x_1^0, x_2^0) t.
\eea These functions vanish in the points corresponding to the
solutions of the equations of motion, and in particular for all
the $A_t \equiv (x_1=0, x_2=0, t>0; \ x_1^0=0, x_2^0=0)$. Labeling
with $\partial_1$ and $\partial_2$ the partial derivatives with
respect to $x_1^0$ and $x_2^0$, the Dini prescription tells us
that if the determinant of the Hessian matrix \be \label{hessiano}
    \frac{\partial (F, G)}{\partial (x_1^0, x_2^0)} =
    \left|
        \begin{array}{cc}
            \partial_1 F & \partial_2 F \\
            \partial_1 G & \partial_2 G
        \end{array}
    \right|
\ee is different from zero in a neighborhood of $A_t$, then we can
explicitate $x_1^0$ and $x_2^0$ as functions of $x_1$, $x_2$ and
$t$, in such neighborhood. This means that the initial point and
the velocities, which depend on it, are univocally determined by
$x_1$ , $x_2$ and $t$ via the equations of motion.

Calculating the determinant we find
\bea
    \frac{\partial (F, G)}{\partial (x_1^0, x_2^0)}
    & = &
    (-1 + \partial_1 \bar{q}_1 t) (-1 + \partial_2 \bar{q}_2 t - \partial_2 \bar{q}_1 t)
    - ( \partial_2 \bar{q}_1 t)(+ \partial_1 \bar{q}_2 t - \partial_1 \bar{q}_1 t)=  \nonumber \\
    \label{hessiano1}
    & = &
    1 + (\partial_2 \bar{q}_1 - \partial_1 \bar{q}_1 - \partial_2 \bar{q}_2) t
    + ( \partial_1 \bar{q}_1 \partial_2 \bar{q}_2 - \partial_2 \bar{q}_1 \partial_1 \bar{q}_2 ) t^2.
\eea so we should ask, for all $x_1^0>0$ and $x_2^0>0$ \be
    \Delta \equiv (\partial_2 \bar{q}_1 - \partial_1 \bar{q}_1 - \partial_2 \bar{q}_2)^2
    - 4 ( \partial_1 \bar{q}_1 \partial_2 \bar{q}_2 - \partial_2 \bar{q}_1 \partial_1 \bar{q}_2 )
\ee to be negative, or in case $\Delta \geq 0$, the zeros of \be
    t_{\pm} = \frac{-\partial_2 \bar{q}_1 + \partial_1 \bar{q}_1 + \partial_2 \bar{q}_2 \pm \sqrt{\Delta}}
                                    {2( \partial_1 \bar{q}_1 \partial_2 \bar{q}_2 - \partial_2 \bar{q}_1 \partial_1 \bar{q}_2 )}
\ee
correspond to non-invertibility points.

The expression we obtain for the determinant is quite untractable,
however we can show uniqueness in a neighborhood of the initial
point $x_1^0=0$, $x_2^0=0$. The motion starting from this point
has zero velocity, and we saw that it gives the high temperature
solution for the mean field spin glass model: Remembering that the
transition to the low temperature is continuous, we can expand the
Hessian for small values of $x_1^0$ and $x_2^0$ and observe that,
for $x_1=0$, $x_2=0$, $t=\beta^2$, the equations of motions become
\bea
    x_1^0 & = & \beta^2 \bar{q}_1 \\
    x_2^0 & = & \beta^2(\bar{q}_2 - \bar{q}_1).
\eea When $x_1^0 \rightarrow 0 $ and $x_2^0 \rightarrow 0$ we have
also $\bar{q}_1 \rightarrow 0$ and $\bar{q}_2 \rightarrow 0$, so
we have an expansion close to the critical point (which is the
only region where the control of the unstable $1$-RSB solution
makes sense for the SK, being the latter $\infty$-RSB).

For $ \bar{q}_1$ and $ \bar{q}_2 $ we have, retaining terms until
the second order: \bea \label{q1approx}
    \bar{q}_1(x_1^0, x_2^0) & \approx & x_1^0 - 2(1-m) x_1^0 x_2^0 - 2(x_1^0)^2 \\
\label{q2approx}
    \bar{q}_2(x_1^0, x_2^0) & \approx & x_1^0 + x_2^0 + m x_2^0 (x_2^0 + 2x_1^0)
\eea
and consequently
\bea
\label{d1q1approx}
    \partial_1 \bar{q}_1(x_1^0, x_2^0) & \approx & 1 - 2(1-m) x_2^0 - 4 x_1^0 \\
\label{d2q1approx}
    \partial_2\bar{q}_1(x_1^0, x_2^0) & \approx & - 2(1-m) x_1^0 \\
\label{d1q2approx}
    \partial_1\bar{q}_2(x_1^0, x_2^0) & \approx & 1 + 2 m x_2^0 \\
\label{d2q2approx}
    \partial_2\bar{q}_2(x_1^0, x_2^0) & \approx & 1 + 2 m x_1^0 + 2 m x_2^0.
\eea
Substituting in (\ref{hessiano1}) we find
\bea
    \frac{\partial (F, G)}{\partial (x_1^0, x_2^0)}
    & \approx &
    1 - 2 \left[1 - x_1^0 - (1-2m)x_2^0 \right] t \nonumber \\
    && +\left[1 - (2-m)x_1^0 - 2(1-2m)x_2^0 + 2m(m-4)x_1^0 x_2^0 + \right.\nonumber \\
    &&  \left. - 8m (x_1^0)^2 - 4m(1-m) (x_2^0)^2 \right] t^2.
\eea and, for $x_1^0, x_2^0 = 0$ (which corresponds to expand the
velocities up to the first order in $x_1^0$ and $x_2^0$) we simply
obtain \be
    \frac{\partial (F, G)}{\partial (x_1^0, x_2^0)} \approx (1-t)^2.
\ee This means that in a neighborhood of $x_1 = x_2 = 0$ we have
uniqueness, provided that we are not exactly at the critical point
$t= \beta^2 = 1 $.

\section{Outlooks and conclusions}

In this paper we enlarged the previously investigated
Hamilton-Jacobi structure for free energy in thermodynamics of
complex systems (tested on the paradigmatic SK model) by merging
this approach with the Broken Replica Symmetry Bound technique. At
the mathematical level the main achievement is the development of
a new method which is autonomously able to give the various steps
of replica symmetry breaking (of the replica trick counterpart).
At a physical level this methods clearly highlights why increasing
the steps of RSB improves the obtained thermodynamics mirroring
these increments in diminishing the approximation of a free field
propagation in an Euclidean space time of an enlarged free energy,
which recovers the proper one of statistical mechanics as a
particular, well defined, limit: the main achievement is paving an
alternative way to understand RSB phenomenon.
\newline
However, when increasing the steps of RSB (making smaller the
potential we neglect, and so smaller the error) there is a price
to pay: each step of replica symmetry breaking enlarges by one
dimension the space for the motion of the mechanical action. As a
consequence the full RSB theory should live on an Hilbert space:
this still deserve more analysis, however the method is already
clear and several application may now stem: for example P-spin
above the Gardner critical temperature could be solved exactly as
well as a consistent part of the plethora of models born in
disordered system statistical mechanics whose solutions implies
only one step of RSB.
\newline
We deserve to investigate both the $K\to\infty$ limit to complete
the theory as well as its simpler immediate applications.

\section*{Appendix: Streaming of the interpolating function $\tilde{\alpha}(t,\bold{x})$}

In this section we show in all details how to get the  streaming
of the interpolating function (\ref{alfa})

          The $t$-streaming of the interpolating function
          $\tilde{\alpha}(t,\bold{x})$ is given by the following
          formula:

                \begin{equation}
                \label{dtalfa}
                    \partial_t \tilde{\alpha}_N(\bold{x},t) = \frac{1}{4} \left(1
                    - \sum_{a=0}^{K}(m_{a+1} - m_a) \langle q_{12}^2(\bold{x},t) \rangle_a
                    \right).
                \end{equation}

                To get this result, let us start by
                \begin{equation}
                \label{dtalfa1}
                    \partial_t \tilde{\alpha}_N(\bold{x},t) = \frac{1}{N} E_0 Z_0^{-1}(\bold{x},t) \partial_t
                    Z_0(\bold{x},t),
                \end{equation}
    and, as it is straightforward to show
    that
                \begin{equation}
                \label{dtza}
                    Z_a^{-1}(\bold{x},t)\partial_t Z_a(\bold{x},t) = E_{a+1} \left( f_{a+1} Z_{a+1}^{-1}(\bold{x},t)\partial_t Z_{a+1}(\bold{x},t)
                    \right),
                \end{equation}
by iteration we get
                \begin{equation}
                \label{dtz0}
                    Z_0^{-1}(\bold{x},t) \partial_t Z_0(\bold{x},t) = E_1 \dots E_K (f_1 \dots f_K Z_K^{-1}(\bold{x},t)\partial_t Z_K(\bold{x},t)).
                \end{equation}
The $t$-derivative of $Z_K$ is then given by
                \begin{equation}
                \label{dtzk}
                    Z_K^{-1}(\bold{x},t) \partial_t Z_K(\bold{x},t) = \frac{1}{4 \sqrt{tN}}
                    \sum_{ij} J_{ij} \omega (\sigma_i \sigma_j),
                \end{equation}
                from which
                \begin{equation}
                \label{dtz01}
                    Z_0^{-1}(\bold{x},t) \partial_t Z_0(\bold{x},t) = \frac{1}{4 \sqrt{tN}}
                    \sum_{ij} E \left( f_1 \dots f_K J_{ij} \omega (\sigma_i \sigma_j)
                    \right),
                \end{equation}
                where we labeled with $E$ the
                global average overall the random variables as
                there is no danger of confusion.
                All the terms into the sum can be worked out
                integrating by parts:
                \begin{eqnarray}
                \label{gaussparti}
                    E \left( f_1 \dots f_K J_{ij} \omega (\sigma_i \sigma_j) \right)
                    & = &
                    \sum_{a=1}^K E \left(
                    f_1 \dots \partial_{J_{ij}}f_a \dots f_K  \omega (\sigma_i \sigma_j) \right) \nonumber \\
                     \quad & + & E \left( f_1 \dots f_K  \partial_{J_{ij}} \omega (\sigma_i \sigma_j) \right).
                \end{eqnarray}
So we need to calculate the explicit expression of the derivatives
with respect to $J_{ij}$ of both $f_a$ as well as $\omega
(\sigma_i \sigma_j)$. For the latter, it is easy to check that
                \begin{equation}
                \label{djijomega}
                    \partial_{J_{ij}} \omega (\sigma_i \sigma_j)
                    = \sqrt{\frac{t}{N}} \left( 1 - \omega^2 (\sigma_i \sigma_j)
                    \right),
                \end{equation}
while for the $f_a$'s we have
                \begin{equation}
                    \partial_{J_{ij}} f_a = m_a f_a \left(Z_a^{-1}(\bold{x},t) \partial_{J_{ij}} Z_a(\bold{x},t) \right)
                    - m_a f_a E_a f_a \left(Z_a^{-1}(\bold{x},t) \partial_{J_{ij}} Z_a(\bold{x},t)
                    \right).
                \end{equation}
By using the analogy of (\ref{dtza}) we get
                \begin{eqnarray}
                    Z_a^{-1}(\bold{x},t) \partial_{J_{ij}} Z_a(\bold{x},t)
                    & = & E_{a+1} \dots E_K \left( f_{a+1} \dots f_K Z_K^{-1} \partial_{J_{ij}} Z_K\right) \nonumber \\
                    \label{djijza}
                    & = & \sqrt{\frac{t}{N}} \tilde{\omega}_a (\sigma_i
                    \sigma_j),
                \end{eqnarray}
                such that
                \begin{equation}
                \label{djijfa}
                    \partial_{J_{ij}} f_a = m_a f_a \sqrt{\frac{t}{N}}
                    \left(\tilde{\omega}_a (\sigma_i \sigma_j) - \tilde{\omega}_{a-1} (\sigma_i \sigma_j) \right).
                \end{equation}
                Substituting (\ref{djijomega}) and (\ref{djijfa}) into (\ref{gaussparti})
                we obtain
                \begin{eqnarray}
                    E \left( f_1 \dots f_K J_{ij} \omega (\sigma_i \sigma_j) \right)
                    & = & \sqrt{\frac{t}{N}} \sum_{a=1}^K m_a \left[ E \left(f_1 \dots f_a
                    \tilde{\omega}_a (\sigma_i \sigma_j) \dots f_K \omega (\sigma_i \sigma_j) \right) \right. \nonumber \\
                    & - & \left. E \left(f_1 \dots f_{a-1} \tilde{\omega}_{a-1} (\sigma_i \sigma_j)
                    \dots f_K \omega (\sigma_i \sigma_j) \right) \right] \nonumber \\
                    & + & \sqrt{\frac{t}{N}} E \left( f_1 \dots f_K (1 - \omega^2 (\sigma_i \sigma_j)) \right).
                \end{eqnarray}
                Overall, an explicit expression for the eq. (\ref{dtalfa1}) is given by
                \begin{eqnarray}
                    \partial_t \tilde{\alpha} & = & \frac{1}{4N^2} \sum_{a=1}^K \sum_{ij} m_a \left[
                    E_0 \dots E_a f_1 \dots f_a \tilde{\omega}_a( \sigma_i \sigma_j)
                    E_{a+1} \dots E_K f_{a+1} \dots f_K \omega( \sigma_i \sigma_j) \right. \nonumber \\
                    & - & \left. E_0 \dots E_{a-1} f_1 \dots f_{a-1} \tilde{\omega}_{a-1}( \sigma_i \sigma_j)
                    E_{a} \dots E_K f_{a} \dots f_K \omega( \sigma_i \sigma_j) \right] \nonumber \\
                    &
                    \label{dtalfa2}
                    + & \frac{1}{4N^2} E f_1 \dots f_K \sum_{ij} (1 - \omega^2(\sigma_i \sigma_j)).
                \end{eqnarray}
               Once introduced the overlap, we can write the
               result:
                \begin{eqnarray}
                    \partial_t \tilde{\alpha} & = & \frac{1}{4} \sum_{a=1}^K m_a
                    (\langle q_{12}^2 \rangle_a - \langle q_{12}^2 \rangle_{a-1})
                    + \frac{1}{4}(1 - \langle q_{12}^2 \rangle_K) \nonumber \\
                    && = \frac{1}{4}\left( \sum_{a=1}^K m_a \langle q_{12}^2 \rangle_a
                    - \sum_{a=0}^K m_{a+1} \langle q_{12}^2 \rangle_a
                    + m_{K+1}\langle q_{12}^2 \rangle_K + 1 - \langle q_{12}^2 \rangle_K\right) \nonumber \\
                    && = \frac{1}{4} \left(1
                    - \sum_{a=0}^{K}(m_{a+1} - m_a) \langle q_{12}^2 \rangle_a \right).
                \end{eqnarray}

Now let us focus on the  $x$-streaming of the interpolating
function  $\tilde{\alpha}(t,\bold{x})$ and show that it is given
by the following formula:

                \begin{equation}
                \label{daalfa}
                    \partial_a \tilde{\alpha}_N(\bold{x},t) = \frac{1}{2} \left(1
                    - \sum_{b=a}^{K}(m_{b+1} - m_b) \langle q_{12}(\bold{x},t) \rangle_b
                    \right).
                \end{equation}
              In analogy with the $t$-streaming we have
                \begin{eqnarray}
                \label{daalfa1}
                    \partial_a \tilde{\alpha}_N(\bold{x},t) & = & \frac{1}{N} E_0 Z_0^{-1}(\bold{x},t) \partial_a Z_0(\bold{x},t), \\
                \label{daza}
                    Z_b^{-1}(\bold{x},t)\partial_a Z_b(\bold{x},t) & = & E_{b+1} \left( f_{b+1} Z_{b+1}^{-1}(\bold{x},t)\partial_a Z_{b+1}(\bold{x},t) \right), \\
                \label{daz0}
                    \Rightarrow Z_0^{-1}(\bold{x},t) \partial_a Z_0(\bold{x},t) & = & E_1 \dots E_K (f_1 \dots f_K Z_K^{-1}(\bold{x},t)\partial_a Z_K(\bold{x},t)), \\
                \label{dazk}
                    Z_K^{-1}(\bold{x},t) \partial_a Z_K(\bold{x},t) & = & \frac{1}{2 \sqrt{x_a}}
                    \sum_{i} J_{i}^a \tilde{\omega} (\sigma_i),
                \end{eqnarray}
              by which
                \begin{equation}
                \label{daalfa2}
                    \partial_a \tilde{\alpha} = \frac{1}{N} \frac{1}{2\sqrt{x_a}}
                    \sum_{i} E \left( f_1 \dots f_K J_{i}^a \tilde{\omega} (\sigma_i) \right).
                \end{equation}
Again by integrating by parts we have
                \begin{equation}
                \label{gausspartia}
                    \partial_a \tilde{\alpha}  =  \frac{1}{N} \frac{1}{2\sqrt{x_a}} \sum_{i=1}^N \left[
                    \sum_{b=1}^K E \left( f_1\dots \partial_{J_i^a}f_b \dots f_K
                    \tilde{\omega}(\sigma_i) \right) \right.
                  \left.
                    + E \left( f_1 \dots f_K  \partial_{J_i^a} \tilde{\omega} (\sigma_i) \right)\right].
                \end{equation}
Let us work out the $J_{i}^a$ by remembering that $Z_b$'s, and
consequently $f_b$'s, do not depend on $J_i^{b+1}, \dots, J_i^K$.
                \begin{equation}
                \label{djiaf}
                    \partial_{J_i^a} f_b =
                    \left\{
                    \begin{array}{ll}
                    0                                                                                                                           & \mbox{if } a > b \\
                    m_a f_a \left(Z_a^{-1}(\bold{x},t) \partial_{J_i^a} Z_a (\bold{x},t) \right)                         & \mbox{if } a = b \\
                    m_b f_b \left(Z_b^{-1}(\bold{x},t) \partial_{J_i^a} Z_b(\bold{x},t) \right)
                    - m_b f_b E_b f_b \left(Z_b^{-1}(\bold{x},t) \partial_{J_i^a} Z_b(\bold{x},t) \right)   & \mbox{if } a < b.
                    \end{array}
                    \right.
                \end{equation}
The same recursion relationship holds in this case as well:
                \begin{equation}
                    Z_b^{-1}(\bold{x},t)\partial_{J_i^a} Z_b(\bold{x},t) =
                    E_{b+1} \dots E_K \left( f_{b+1} \dots f_K Z_{K}^{-1}(\bold{x},t)\partial_{J_i^a} Z_{K}(\bold{x},t)
                    \right).
                \end{equation}
                Furthermore
                \begin{equation}
                    Z_K^{-1}(\bold{x},t) \partial_{J_{i}^a} Z_K (\bold{x},t) =
                  \sqrt{x_a} \tilde{\omega}(\sigma_i),
                \end{equation}
                from which we get
                \begin{equation}
                    Z_b^{-1}(\bold{x},t) \partial_{J_{i}^a} Z_b (\bold{x},t)
                    = \sqrt{x_a} \tilde{\omega}_b (\sigma_i).
                \end{equation}
                Consequently, eq.s (\ref{djiaf}) can be written as
                \begin{equation}
                \label{djiaf1}
                    \partial_{J_i^a} f_b =
                    \left\{
                    \begin{array}{ll}
                    0                                                                                                                           & \mbox{if } a > b \\
                    m_a f_a \sqrt{x_a} \tilde{\omega}_a (\sigma_i)                      & \mbox{if } a = b \\
                     \sqrt{x_a} m_b f_b \left( \tilde{\omega}_b (\sigma_i)
                     -\tilde{\omega}_{b-1} (\sigma_i) \right)                                       & \mbox{if } a < b.
                    \end{array}
                    \right.
                \end{equation}
    The last thing missing is evaluating the derivative of the
    state
                \begin{equation}
                \label{djiaomega}
                    \partial_{J_{i}^a} \omega (\sigma_i)
                    = \sqrt{x_a} \left( 1 - \omega^2 (\sigma_i)
                    \right),
                \end{equation}
                so to write, via the overlap,
                the analogous for the generalized states.
Substituting eq.s (\ref{djiaomega}) and (\ref{djiaf1}), once
expressed via overlaps, into (\ref{gausspartia}) we obtain eq.
(\ref{daalfa}).

\section*{Acknowledgements}
The authors are pleased to thank  Pierluigi Contucci, Giuseppe
Genovese, Sandro Graffi, Isaac Perez-Castillo and Tim Rogers for
useful discussions.
\newline
AB work is supported by the Smart-Life Project grant.

%\end{acknowledgements}

% BibTeX users please use one of
%\bibliographystyle{spbasic}      % basic style, author-year citations
%\bibliographystyle{spmpsci}      % mathematics and physical sciences
%\bibliographystyle{spphys}       % APS-like style for physics
%\bibliography{}   % name your BibTeX data base

\begin{thebibliography}{9}

\bibitem{barra3} A. Agostini, A. Barra, L. De Sanctis
\textit{Positive-Overlap transition and Critical Exponents in mean
field spin glasses}, J. Stat. Mech. P11015 (2006).

\bibitem{ac} M. Aizenman, P. Contucci, {\em On the stability of the
quenched state in mean field spin glass models}, J. Stat. Phys.
{\bf 92}, 765-783 (1998).
%Pagina

\bibitem{alr} M. Aizenman, J. Lebowitz and D. Ruelle, {\em Some rigorous
results on the Sherrington-Kirkpatrick spin glass model}, Commun.
Math. Phys. {\bf 112}, 3-20 (1987).
%Pagina

\bibitem{ass} M. Aizenman, R. Sims, S. L. Starr,
\emph{An Extended Variational Principle for the SK Spin-Glass
Model}, Phys. Rev. B \textbf{68}, 214403 (2003).

\bibitem{bara} R. Albert, A. L. Barabasi
{\em Statistical mechanics of complex networks}, Rev.  Mod. Phys.
\textbf{74}, 47-97 (2002).
%Pagina

\bibitem{amit} D.J. Amit, {\em Modeling brain function. The world
of attractor neural networks}, Cambridge University Press, New
York (1989).

\bibitem{arguin} L.P. Arguin, M. Aizenman,  {\em On the Structure of Quasi-Stationary Competing Particle
Systems}, Ann. Probab. \textbf{37}, 3, 1080-1113 (2009).
%aggiunte pagine

\bibitem{arguin2} L.P. Arguin, {\em Spin-glass computation and probability
cascades}, J. Stat. Phys. \textbf{126}, 951-976 (2007).

\bibitem{barra1} A. Barra,
\textit{Irreducible free energy expansion and overlap locking in
mean field spin glasses}, J. Stat. Phys. \textbf{123}, 601-614
(2006).

\bibitem{barra2} A. Barra,
\textit{The mean field Ising model trough interpolating
techniques}, J. Stat. Phys. \textbf{132}, 18-32 (2008).

\bibitem{bouchaud} J.P. Bouchaud, M. Potters, {\em Theory of financial risk and derivative
pricing. From statistical physics to risk management}, Cambridge
Univ. Press (2000).
%questo è un libro non so se va bene scritto così

\bibitem{bovierbook} A. Bovier, {\em Statistical Mechanics of Disordered
Systems. A Mathematical Perspective}, Cambridge Series \textbf{18}
(2006).
%questo è un libro non so se va bene scritto così

\bibitem{bovier} A. Bovier, I. Kurkova, {\em
Local statistics of energy levels in spin glasses}, J. Stat. Phys.
\textbf{126}, 933-949 (2007).
%pagine

\bibitem{CG} W. Brock, S. Durlauf, {\em Discrete choice with social
interactions}, Review of Economic Studies \textbf{68}, 235-260
(2001).
%aggiunte pagine

\bibitem{carmona} P. Carmona, Y.  Hu, {\em Universality in
Sherrington Kirkpatrick spin glass model}, Ann. Inst. H. Poin.
(B), Prob. et Stat. \textbf{42}, Issue 2, 215-222 (2006).
%aggiunte pagine e issue

\bibitem{comets} F. Comets, J. Neveu, {\em The Sherrington-Kirkpatrick
model of spin glasses and stochastic calculus: the high
temperature case}, Commun. Math. Phys. {\bf 166}, n. 3, 549-564
(1995).
%aggiunte pagine e numero

\bibitem{contucci} P. Contucci, C. Giardin\`a,
{\em Spin Glass Stochastic Stability: A rigorous proof}, Annales
Henri Poincar\`e, \textbf{6},  915 - 923 (2005).
%aggiunte pagine

\bibitem{hertz} K. H. Fischer, J. A. Hertz, {\em Spin Glasses},
Cambridge Studies in Magnetism (1993).

\bibitem{io1} G. Genovese, A. Barra,
{\em A mechanical approach to mean field spin models}, J. Math.
Phys. \textbf{50} (2009).
%non so se va aggiunto qualcosa

\bibitem{sum-rules} F. Guerra, {\em Sum rules for the free energy in the mean
field spin glass model}, in {\em Mathematical Physics in
Mathematics and Physics: Quantum and Operator Algebraic Aspects},
Fields Institute Communications {\bf 30}, 161-170 (2001).
%aggiunte pagine

\bibitem{g3} F. Guerra, \emph{Broken Replica Symmetry Bounds
in the Mean Field Spin Glass Model}, Commun. Math. Phys.
\textbf{233:1}, 1-12 (2003)


\bibitem{limterm} F. Guerra, F. L. Toninelli, {\em
The Thermodynamic Limit in Mean Field Spin Glass Models}, Commun.
Math. Phys. {\bf 230:1}, 71-79 (2002).


\bibitem{CLT} F. Guerra, F. L. Toninelli, {\em Central limit theorem for
fluctuations in the high temperature region of the
Sherrington-Kirkpatrick spin glass model}, J. Math. Phys.
\textbf{43}, 6224-6237 (2002).


\bibitem{quadratic} F. Guerra, F. L. Toninelli, {\em Quadratic replica coupling for
the Sherrington-Kirkpatrick mean field spin glass model}, J. Math.
Phys. {\bf 43}, 3704-3716 (2002).
%aggiunte pagine

\bibitem{gg} S. Ghirlanda, F. Guerra, {\em General properties of overlap
distributions in disordered spin systems. Towards Parisi
ultrametricity}, J. Phys. A {\bf 31}, 9149-9155 (1998).

\bibitem{sk} S. Kirkpatrick, D. Sherrington, {\em Solvable model of
a spin-glass}, Phys. Rev. Lett. {\bf 35}, 1792-1796 (1975).
%pagina

\bibitem{sk2} S. Kirkpatrick, D. Sherrington, {\em Infinite-ranged models
of spin-glasses}, Phys. Rev. B {\bf 17}, 4384-4403 (1978).
%pagina

\bibitem{MPV} M. Mezard, G. Parisi, M.A. Virasoro, {\em Spin glass theory and
beyond}, World Scientific, Singapore (1987).

\bibitem{science} M. M\'ezard, G. Parisi, R. Zecchina,
{\em Analytic and Algorithmic Solution of Random Satisfiability
Problems}, Science {\bf 297}, 812-815 (2002).

\bibitem{panchenko} D. Panchenko, {em A connection between Ghirlanda-Guerra identities and
ultrametricity}, arXiv:0810.0743, (2008).

\bibitem{parisi} G. Parisi, {\em A simple model for the immune
network}, P.N.A.S. vol. 87, no. \textbf{1}, 429-433 (1990).
%aggiunto volume e pagine


\bibitem{parisi2} G. Parisi, {\em Toward a mean field theory for spin glasses},
Phys. Lett. A {\bf 73}, 3, 203-205 (1979).
%aggiunta issue (3) e pagina finale


\bibitem{parisi3} G. Parisi, {\em A sequence of approximated solutions to the
S-K model for spin glasses}, J. Phys. A: Math. Gen. {\bf 13},
L-115 (1980).
%aggiunto math. gen.

\bibitem{parisi4} G. Parisi, {\em The order parameter for spin glasses: a
function on the interval $0-1$}, J. Phys. A: Math. Gen. {\bf 13},
1101-1112 (1980).
%aggiunto math. gen. e pagina finale

\bibitem{pastur} L. Pastur, M. Scherbina, {\em The absence of self-averaging
of the order parameter in the Sherrington-Kirkpatrick model}, J.
Stat. Phys. {\bf 62},  Nos 1/2, 1-19 (1991).
%aggiunto Nos e pagine


\bibitem{talaRSB} M. Talagrand, {\em Replica symmetry breaking and exponential
inequalities for the Sherrington Kirkpatrick model}, Ann. Probab.
{\bf 28}, no. 3, 1018-1062 (2000).
%pagina finale e no. 3

\bibitem{talaHT} M. Talagrand, {\em On the high temperature region of the
Sherrington-Kirkpatrick model}, Ann. Probab. \textbf{30}, no.1,
364-381 (2002).
%aggiunto numero

\bibitem{t4} M. Talagrand,
\emph{The Parisi Formula}, Annals of Mathematics \textbf{163}, Vol
1, 221-263 (2006).

\bibitem{challenge} M. Talagrand, {\em The Sherrington Kirkpatrick model:
a challenge for mathematicians}, Probab. Rel. Fields {\bf 110},
109-176 (1998).


\end{thebibliography}

% Non-BibTeX users please use

\end{document}